\documentclass[%
 reprint,
 amsmath,amssymb,
 aps,
prb,
]{revtex4-2}

\usepackage{graphicx}
\usepackage{dcolumn}
\usepackage{bm}

\begin{document}


\title{Dispersive-induced magnon blockade with a superconducting qubit}

\author{Zeng-Xing Liu$^{1}$}\email{liuzx@dgut.edu.cn}
\author{Yan-Hua Wu$^{1}$}
\author{Jing-Hua Sun$^{1}$$^{2}$}


\affiliation{$^{1}$ School of Electrical Engineering $\&$ Intelligentization, Dongguan University of Technology, Dongguan, 523808, China}
\affiliation{$^{2}$ Quantum Science Center of Guangdong-Hong Kong-Macao Greater Bay Area, Shenzhen-Hong Kong International Science and Technology Park, No. 3 Binglang Road, Futian District, Shenzhen, Guangdong 518045, China}


\date{\today}

\begin{abstract}
We investigate the magnon blockade effect in a quantum magnonic system operating in the strong dispersive regime, where a superconducting qubit interacts dispersively with a magnonic mode in a yttrium-iron-garnet sphere.
By solving the quantum master equation, we demonstrate that the magnon blockade, characterized by the second-order correlation function $g^{(2)}(0) \rightarrow 0.04$, emerges under optimal dispersive coupling and driving detuning.
The mechanism is attributed to suppressed two-magnon transitions as a result of qubit-induced anharmonicity. Notably, our study identifies the critical role of dispersive interaction strength and environmental temperature, showing that magnon blockade remains observable under experimentally achievable cryogenic conditions.
This work extends the magnon blockade effect into the dispersive regime, offering a robust platform for single-magnon manipulation and advancing applications in quantum sensing and information processing.
\end{abstract}

\maketitle

\section{INTRODUCTION}

Quantum magnonics, as a highly interdisciplinary field, combines spintronics, quantum optics, and quantum information science, making the magnonic system a promising platform for studying quantum physics, but also significantly expanding the horizon of traditional spintronics \cite{mode2,Hybrid,quantum1,quantum2,quantum3,FOR3,Xiong}.
Many intriguing quantum effects of magnons have been discovered, both theoretically and experimentally, ranging from magnon-photon entanglement \cite{entanglement1,entanglement2,entanglement3,entanglement4,entanglement5,FOR1} and squeezed states of magnons \cite{Squeezed,Squeezed1}, to the generation of a magnonic Schr\"{o}dinger cat state \cite{cat,cat1,cat2,cat3,cat4} and quantum many-body states including quasi-equilibrium magnon Bose-Einstein condensation \cite{Einstein,Einstein1,Einstein2,Einstein3}.
In recent years, a promising way to observe quantum effects of the magnon through the coupling between magnetostatic modes and superconducting qubits has been extensively studied and progressed enormously \cite{qubit1,qubit3,qubit4,qubit5,Absorption}.
Of particular interest is that it provides a promising architecture for the development of quantum sensors based on magnonic systems.
For example, single-magnon detection in a superconducting-qubit-based hybrid quantum system in the strong dispersive regime has been experimentally demonstrated  \cite{qubit4,qubit5}, which establishes the single-photon detector counterpart \cite{non-demolition} to the emerging field of magnonics.

It is well known that one of the most important aspects in quantum magnonics is the implementation of magnon control at a single-magnon level, which has significant practical relevance for quantum sensors, quantum information processing, and magnonic quantum simulation schemes \cite{network1,network2,network,network6,network66,network666,network3,network4,network5}.
Magnon blockade, a direct analog of Coulomb blockade in condensed matter physics \cite{Coulomb1,Coulomb2}, provides the possibility of engineering single-magnon-level quantum manipulation and has attracted a great deal of attention \cite{blockade1,blockade2,blockade3,blockade4,blockade5,blockade6,FOR2,blockade7,blockade8,blockade9,Wang11,blockade10,blockade11}.
Recently, the magnon blockade effect was demonstrated in a ferromagnet-superconductor quantum system working in the resonant strong coupling regime, where the energy level diagram of the magnon can be changed to an anharmonic form by the coherent coupling between the qubit and the magnon, thus achieving magnon blockade \cite{blockade1}.
However, the investigation of magnon blockade effect in a hybrid magnon-superconducting qubit quantum system operating in the strong dispersive regime is still new and remarkable for quantum magnonics.
Furthermore, the dispersive interaction between the magnonic and the qubit mode is of particular interest because it has important applications in quantum non-demolition measurement of magnon number, as the experiments demonstrated in Refs \cite{qubit4,qubit5}.
Therefore, investigation of magnon blockade in the strong dispersive regime is very necessary in both fundamental science and technical application.

Here, we extend the study of the magnon blockade effect in quantum magnonics under a strong dispersive regime.
By numerically solving the quantum master equation, we elucidate the dependence of the second-order magnon coherence function $g^{(2)}(0)$ on dispersive interaction and drive detuning, identifying optimal conditions for the emergence of magnon blockade.
Furthermore, we provide a clear explanation for the generation process of the magnon blockade in quantum magnonics through the analysis of magnon resonant transitions.
Examining the impact of environmental temperature on $g^{(2)}(0)$ supports the observability of the magnon blockade effect under current experimental conditions \cite{qubit5}.  
\begin{figure*}[htbp]
\centering
\includegraphics [width=0.75\linewidth] {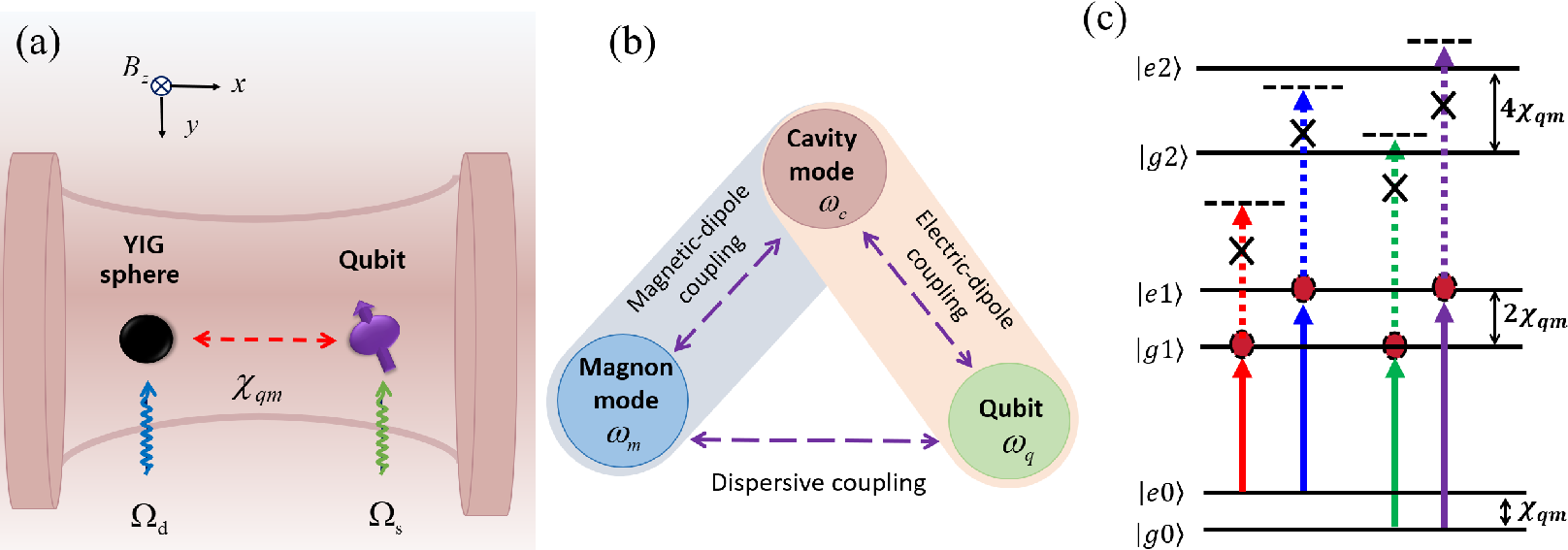}
\caption{(a) The quantum magnonic system consists of a superconducting qubit and a single crystalline YIG sphere inside a three-dimensional microwave cavity.
A uniform magnetic field (with the strength $B$) is applied along the z direction to saturate the magnetization.
Both the Kittel mode and the qubit are controlled by time-dependent driving fields with the driving amplitude $\Omega_{d}$ and $\Omega_{s}$, respectively.
(b) The Kittel mode (with frequency $\omega_{m}$) and the qubit ($\omega_{q}$) are coupled to the microwave cavity mode ($\omega_{c}$) through magnetic-dipole and electric-dipole couplings, while the Kittel mode couples dispersively to the qubit with the dispersive coupling strength $\chi_{qm}$.
(c) Diagram of the energy levels of the dispersive magnon-qubit system with $|g\rangle$ and $|e\rangle$ being the ground and excited states of the qubit, and $|n\rangle(n=0, 1, 2)$ being the magnon number states.}
\label{fig:1}
\end{figure*}
Thus, our results will help to explore the undiscovered magnon traits within the realm of quantum magnonics and may find applications in designing single magnon emitters, which are crucial for quantum control of magnons \cite{network2,network3}.

\section{Physical model and theory}
The quantum magnonic system is shown schematically in Fig. \ref{fig:1}(a), where a Transmon-type superconducting qubit and a single-crystal yttrium-iron-garnet (YIG) sphere are placed inside a three-dimensional microwave copper cavity.
A uniform magnetic field is applied along the z-direction to the YIG sphere, magnetizing it to saturation.
By applying a microwave drive field, the magnetostatic mode (or Kittel mode) of the YIG sphere can be excited \cite{magnon}.
The resonance frequency of the Kittel mode is directly proportional to the amplitude of the uniform magnetic field, i.e., $\omega_{m} = \gamma B$, where $\gamma/2\pi = 28 \rm{GHz/T}$ is the gyromagnetic ratio.
In experiments, the amplitude of the applied magnetic field can be tuned by changing the current in the coil, thereby dynamically adjusting the resonance frequency of the Kittel mode \cite{YIG}.
The hybrid system hosts three modes of interest, i.e., the Kittel mode in the YIG sphere (with tunable frequency $\omega_{m}$), the qubit (with frequency $\omega_{q}$), and the microwave cavity mode (with frequency $\omega_{c}$).
Among them, the microwave cavity mode interact with the qubit through an electric-dipole interaction, as shown in Fig. \ref{fig:1}(b).
Under the rotating wave approximation, the interaction is described by the Jaynes-Cummings Hamiltonian with \cite{ac-Stark} (we set $\hbar$ = 1 and apply it to the full text)
\begin{eqnarray}
  H_{cq} = g_{cq}(c^{\dagger}\sigma_{-}+c\sigma_{+}),
\end{eqnarray}
where $c (c^{\dagger})$ is the annihilation (creation) operator of the cavity mode and $g_{cq}$ is the coupling strength between the cavity mode and the qubit.
$\sigma_{+}=|e\rangle\langle g|$ ($\sigma_{-}=|g\rangle\langle e|$) is the raising (lowering) operator of the qubit with $|g\rangle$ the ground state and $|e\rangle$ the excited state.
Likewise, the Kittel mode couples to the cavity mode through a magnetic-dipole interaction with \cite{quantum2}
\begin{eqnarray}
  H_{cm} = g_{cm}(c^{\dagger}m+cm^{\dagger}),
\end{eqnarray}
where $m (m^{\dagger})$ is the annihilation (creation) operator of the Kittel mode, and $g_{cm}$ is the coupling strength between the cavity mode and the Kittel mode.

In particular, with the microwave cavity mode far-detuned from the qubit and the Kittel mode, i.e., $|\omega_{c}-\omega_{q}|\gg g_{cq}, g_{cm}$, the microwave cavity mode is adiabatically eliminated \cite{mode2}.
Moreover, if the resonance frequency of the Kittel mode is tuned to be nearly resonant with that of the qubit, i.e., $|\omega_{q}-\omega_{m}|\ll g_{cq}, g_{cm}$, the interaction between the qubit and the Kittel mode is described with \cite{qubit5}
\begin{eqnarray}\label{equ:12}
  H_{qm} = g_{qm}(m^{\dagger}\sigma_{-}+m\sigma_{+}),
\end{eqnarray}
where $g_{qm}$ is the coupling strength between the qubit and the Kittel mode.
The resonant interaction between the qubit and the Kittel mode is a cavity-mediated interaction and is the building block of quantum magnonics \cite{qubit1,qubit3,qubit4,qubit5}.
Of particular interest is the case when the qubit is strongly detuned from the Kittel mode, i.e., $\Delta_{qm} \equiv |\omega_{q}-\omega_{m}|\gg g_{qm}$, the system enters the dispersive regime \cite{qubit4,qubit5}.
In this dispersive regime, the Hamiltonian of the interaction between the qubit and the magnon mode becomes \cite{ac-Stark,ac-Stark1}
\begin{eqnarray}\label{equ:01}
  H_{disp} &=&
  \frac{1}{2}\left[2\chi_{qm}(m^{\dagger}m+\frac{1}{2})\right]\sigma_{z},
\end{eqnarray}
where $\sigma_{z} = |e\rangle\langle e|-|g\rangle\langle g|$ is the Pauli operator for the qubit and $\chi_{qm}={g_{qm}^{2}}/{\Delta_{qm}}$ as the dispersive coupling strength.
The term $\frac{1}{2}[2\chi_{qm}(m^{\dagger}m+1/2)]$ can be interpreted as the ac-Stark/Lamb shift (also known as the dispersive shift) of the qubit transition frequency.
In other words, the dispersive interaction between the qubit and magnon results in a frequency shift of the qubit by $2\chi_{qm}$ for each magnon in the Kittel mode \cite{qubit5}.

To simulate the magnon blockade effect, the YIG sphere is pumped directly with a microwave field $\textbf{h}_{d} = h_{d}\cos(\omega_{d}t)\textbf{e}_{y}$, with frequency $\omega_{d}$ and amplitude $h_{d}$.
The corresponding Hamiltonian is \cite{Kerr}
\begin{eqnarray}\label{equ:011}
  H_{d} = \mu_{0}\int_{V_{m}}\textbf{M}\cdot\textbf{h}_{d}d\tau
  =\Omega_{q}(S^{+}+S^{-})(e^{i\omega_{d}t}+e^{-i\omega_{d}t}),\nonumber \\
\end{eqnarray}
where $\mu_{0}$ is the vacuum permeability and $V_{m}$ is the volume of the YIG sphere.
$\textbf{M}$ is the magnetization of the YIG sphere and $S^{\pm} \equiv S_{x} \pm iS_{y}$ are the macrospin operators.
$\Omega_{q}$ denotes the coupling strength between each single spin and the pumping field.
Using the Holstein-Primakoff transformation \cite{HP}, i.e., $S^{+} = \sqrt{2S-m^{\dagger}m}m$ and $S^{-} = m^{\dagger}\sqrt{2S-m^{\dagger}m}$, the macrospin operators $S^{\pm}$ can be converted to the magnon operators $m(m^{\dagger})$, where $S$ is the spin quantum number of the macrospin.
Furthermore, under a weak drive, the low-lying excitation condition $\langle m^{\dagger}m \rangle/(2S)\ll1$ can easily be satisfied owing to the large number of spins in the YIG sphere \cite{Low}.
Thus, $\sqrt{2S-m^{\dagger}m}$ can be expanded up to the first order of $m^{\dagger}m/4S$, i.e., $\sqrt{2S-m^{\dagger}m} \approx \sqrt{2S}[1-m^{\dagger}m/(4S)]$, and then the Hamiltonian in Eq. (\ref{equ:011}) becomes
\begin{eqnarray}
  H_{d} = \Omega_{d}(m^{\dagger}e^{-i\omega_{d}t} + me^{i\omega_{d}t}).
\end{eqnarray}
Here, the fast oscillating terms are neglected via the rotating-wave approximation. $\Omega_{d} \equiv \sqrt{2S}\Omega_{q}$ is the Rabi frequency of the microwave drive field, and $1-m^{\dagger}m/(4S)\approx1$ is taken.
Also, a control field at frequency $\omega_{s}$ acts on the qubit [shown in Fig. \ref{fig:1}(a)], which can be described by the Hamiltonian as \cite{qubit5}
\begin{eqnarray}
  H_{s} = \Omega_{s}(\sigma_{+}e^{-i\omega_{s}t}+\sigma_{-}e^{i\omega_{s}t}),
\end{eqnarray}
with $\Omega_{s}$ the qubit control strength.
Therefore, the total Hamiltonian of such hybrid system operating in the dispersive regime can be written as
\begin{eqnarray}\label{equ:01}
  H &=& \omega_{m}m^{\dagger}m+
  \frac{1}{2}\left[\omega_{q}+2\chi_{qm}(m^{\dagger}m+\frac{1}{2})\right]\sigma_{z}
  +H_{d}+H_{s}, \nonumber \\
\end{eqnarray}
To make the driving term time independent, the Hamiltonian can be transformed to a doubly-rotating frame with respect to the frequencies of the qubit ($\omega_{s}$) and magnon ($\omega_{d}$) control fields by the unitary transformation, i.e., $\mathbf{U} = \exp[-i\omega_{s}(\sigma_{z}/2)t-i\omega_{d}m^{\dagger}mt]$.
That is \cite{qubit3}
\begin{eqnarray}\label{equ:02}
  H' =&& \mathbf{U}^{\dagger}H\mathbf{U}-i\mathbf{U}^{\dagger}\frac{\partial\mathbf{U}}{\partial t}\nonumber \\
  =&&\Delta_{m}m^{\dagger}m+\frac{1}{2}\Delta_{q}\sigma_{z}+\frac{1}{2}(2\chi_{qm}m^{\dagger}m)\sigma_{z}\nonumber \\
  &&+\Omega_{s}(\sigma_{+}+\sigma_{-})+\Omega_{d}(m^{\dagger}+m),
\end{eqnarray}
where $\Delta_{m} \equiv \omega_{m}-\omega_{d}$ ($\Delta_{q} \equiv \tilde{\omega}_{q}-\omega_{s}$, $\tilde{\omega}_{q} = \omega_{q}+\chi_{qm}$) is the detuning between the Kittel mode (qubit) frequency and the magnon excitation (qubit control) frequency.

In practical situations, coupling to additional uncontrollable bath degrees of freedom leads to energy relaxation and dephasing within the system.
Here we assume that the magnon mode and qubit are connected with two individual vacuum baths. Thus, the dynamics of the system can be described by the quantum master equation \cite{mast1}:
\begin{eqnarray}\label{equ:03}
  \dot{\rho} &=& -i[H', \rho]+\mathcal{D}\rho,
\end{eqnarray}
where $\rho$  is the density matrix in the double-rotating frame.
$\mathcal{D}$ is the dissipator super-operator, which is of the Lindblad form $\mathcal{D}\rho=\sum_{\jmath}\Gamma_{\jmath}\left(\mathcal{L}_{\jmath}\rho\mathcal{L}_{\jmath}^{\dagger}
  -\frac{1}{2}\left\{\mathcal{L}_{\jmath}^{\dagger}\mathcal{L}_{\jmath}, \rho\right\} \right)$, with
$\mathcal{L}_{\jmath}$ a set of quantum jump operators and $\Gamma_{\jmath}$ the rates governing the dissipative dynamics.
The processes, rates, and operators considered in the numerical simulations are given in Table \ref{tab:0}.
\begin{table}[h!]
 	\caption{\label{tab:0}
  The processes, rates, and operators considered in the Lindblad master equation for the numerical simulations.}
  \begin{ruledtabular}
\begin{tabular}{ccc}
 	Process  &  Rate $\Gamma_{\jmath}$ &  Operator $\mathcal{L}_{\jmath}$  \\
\hline
 		Qubit relaxation & $\kappa_{1}(1+n_{th})$  & $\sigma_{-}$ \\
 		Qubit excitation & $\kappa_{1}n_{th}$  &  $\sigma_{+}$\\
 		Qubit pure dephasing & $2\kappa_{\varphi}$  &  $\sigma_{z}$\\
\hline
 		Magnon relaxation & $\kappa_{m}(1+m_{th})$ &  m \\
 		Magnon excitation & $\kappa_{m}m_{th}$  &  $m^{\dagger}$
    \end{tabular}
\end{ruledtabular}
 \end{table}
Here, $\kappa_{m}$ and $\kappa_{q}$ are, respectively, the linewidth of the magnonic mode and the qubit.
The qubit pure dephasing rate $\kappa_{\varphi}=\frac{1}{2}(\kappa_{1}+\kappa_{q})$ is determined from the qubit relaxation rate $\kappa_{1}$ and linewidth $\kappa_{q}$.
$n_{th}$ and $m_{th}$ are the thermal noise numbers of the qubit and magnonic mode, respectively.
The steady-state density-matrix operator $\rho_{ss}$ of the system can be obtained by numerically solving the Lindblad master equation by utilizing QuTiP \cite{QuTiP}.
Thus the magnon-number distributions $P_{n} = {\rm{Tr}}\big[|n\rangle\langle n|\rho_{ss}\big]$, with $|n\rangle(n=0, 1, 2)$ being the magnon number states, and the second-order magnon coherence function $g^{(2)}(0) ={\rm{Tr}}(m^{\dagger}m^{\dagger}mm\rho_{ss})/[{\rm{Tr}}(m^{\dagger}m\rho_{ss})]^{2}$ \cite{quantum1} can be obtained accordingly.
Physically, the second-order magnon coherence function $g^{(2)}(0)$ well describes the spatial and temporal distribution characteristics of magnons.
Specifically, $g^{(2)}(0)$ represents the ratio of the probability of detecting two magnons simultaneously to the square of the probability of detecting a single magnon at zero time delay.
When $g^{(2)}(0)<1$, it indicates that magnons exhibit an antibunching effect, meaning that magnons tend not to appear simultaneously.
In particular, when $g^{(2)}(0)\rightarrow 0$, it implies that the probability of magnons appearing simultaneously is almost zero, similar to the photon blockade effect in quantum optics \cite{block}, and is referred to as the magnon blockade effect \cite{blockade1}.
Quantum magnonics in the dispersive regime play an extremely important role in the quantum control of magnons \cite{qubit4,qubit5}.
However, to date, the magnon blockade effect in the dispersive regime of quantum magnonics remains largely unexplored.

\section{RESULTS AND DISCUSSIONS}
\begin{figure}[htbp]
\centering
\includegraphics [width=1\linewidth] {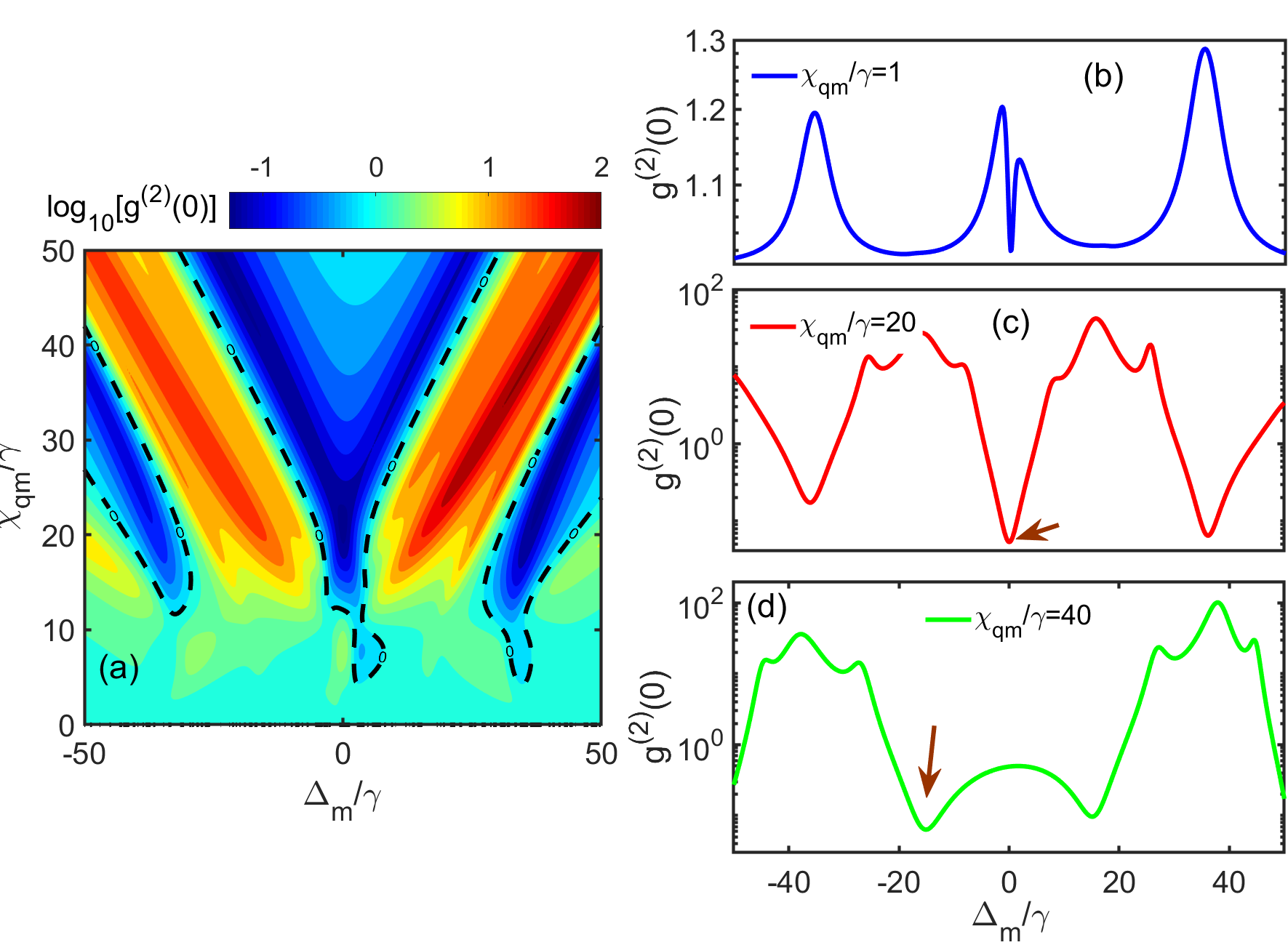}
\caption{(a) The contour plot of the second-order magnon coherence function $\rm{log}_{10}[g^{(2)}(0)]$ varies with the driving detuning $\Delta_{m}$ and the dispersive coupling strength $\chi_{qm}$.
Here, for convenience, $\gamma=2\pi\times1\rm{MHz}$ is used to scale the frequencies throughout the article.
(b)-(d) The second-order magnon coherence function $g^{(2)}(0)$ under the different dispersive coupling strength $\chi_{qm}/\gamma$ = 1, 20, and 40, respectively.
The simulation parameters are chosen from the recent experiment \cite{qubit4,qubit5} $\Omega_{s}/\gamma = 15$, $\Omega_{d}/\gamma = 0.1$, $\Delta_{q}/\gamma = -20$, $\kappa_{m}/\gamma=1.4$, $\kappa_{q}/\gamma=1.2$, and $\kappa_{1}/\gamma=1$. The thermal magnon and qubit number $m_{th} = n_{th} = 0$.}
\label{fig:2}
\end{figure}

\begin{figure}[htbp]
\centering
\includegraphics [width=1\linewidth] {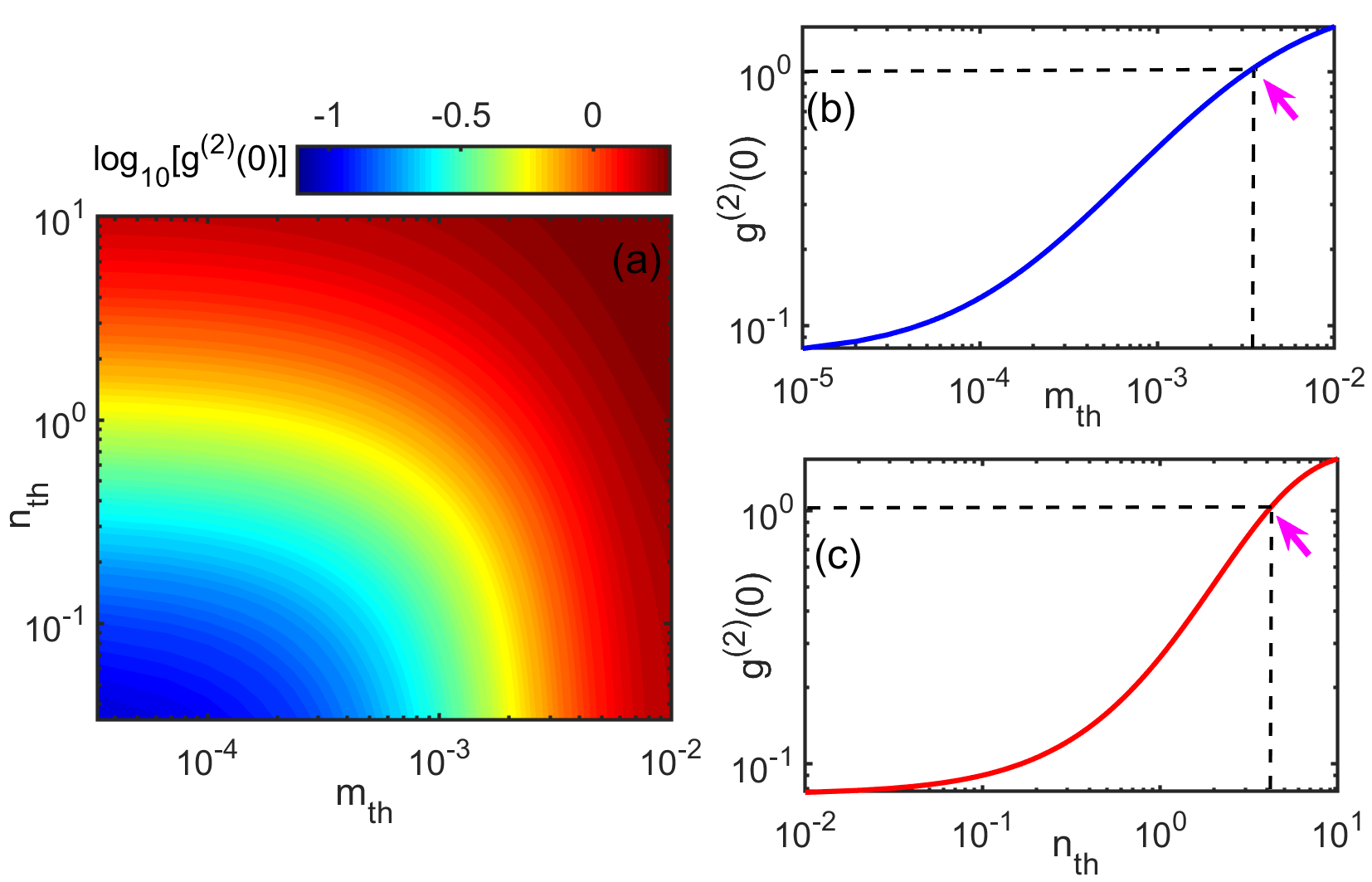}
\caption{(a) The contour plot of the second-order magnon coherence function $\rm{log}_{10}[g^{(2)}(0)]$ varies with the thermal noise numbers of the qubit ($n_{th}$) and magnonic mode ($m_{th}$).
(b) and (c) The second-order magnon coherence function $g^{(2)}(0)$ as a function of the thermal noise numbers $m_{th}$ (with fixed $n_{th}$ = 0) and $n_{th}$ (with fixed $m_{th}$ = 0), respectively.
We set the dispersive coupling strength $\chi_{qm}/\gamma$ = 20, the driving detuning $\Delta_{m}/\gamma$ = 0, and the other parameters are the same as those in Fig. \ref{fig:2}.}
\label{fig:5}
\end{figure}

To study the magnon blockade effect in quantum magnonics within the dispersive regime, the second-order magnon coherence function $g^{(2)}(0)$ varies with the driving detuning $\Delta_{m}$ and the dispersive coupling strength $\chi_{qm}$ is plotted in Fig. \ref{fig:2}(a).
The black contour lines mark the value of the second-order magnon coherence function $\rm{log}_{10}[g^{(2)}(0)] = 0$ [i.e., $g^{(2)}(0) = 1$], which is a clear boundary between the two different statistical properties that the magnonic mode satisfies \cite{mast1}.
To be specific, outside the region bounded by the black contour lines, the second-order magnon coherence function $g^{(2)}(0)>1$ indicates that the steady-state magnon number follows a super-Poissonian distribution, and in this case, the magnonic mode exhibits the bunching effect.
Of interest is that within the contour, the second-order magnon coherence function $0<g^{(2)}(0)<1$ reveals that the magnonic mode exhibits anti-bunching behavior (the steady-state magnon number follows a sub-Poissonian distribution) \cite{blockade2}, which is a quantum-mechanical property of the magnon, revealing the particle-like behavior of the magnonic mode.
Apart from that, it can be observed that the dispersive coupling strength plays a crucial role in the regulation of the quantum properties of the magnons.
In the concerned weak dispersive coupling regime, i.e., $\chi_{qm}/\gamma=1$, as shown in Figs. \ref{fig:2}(b), the magnon number satisfies the Poisson distribution, 
\begin{figure}[htbp]
\centering
\includegraphics [width=1\linewidth] {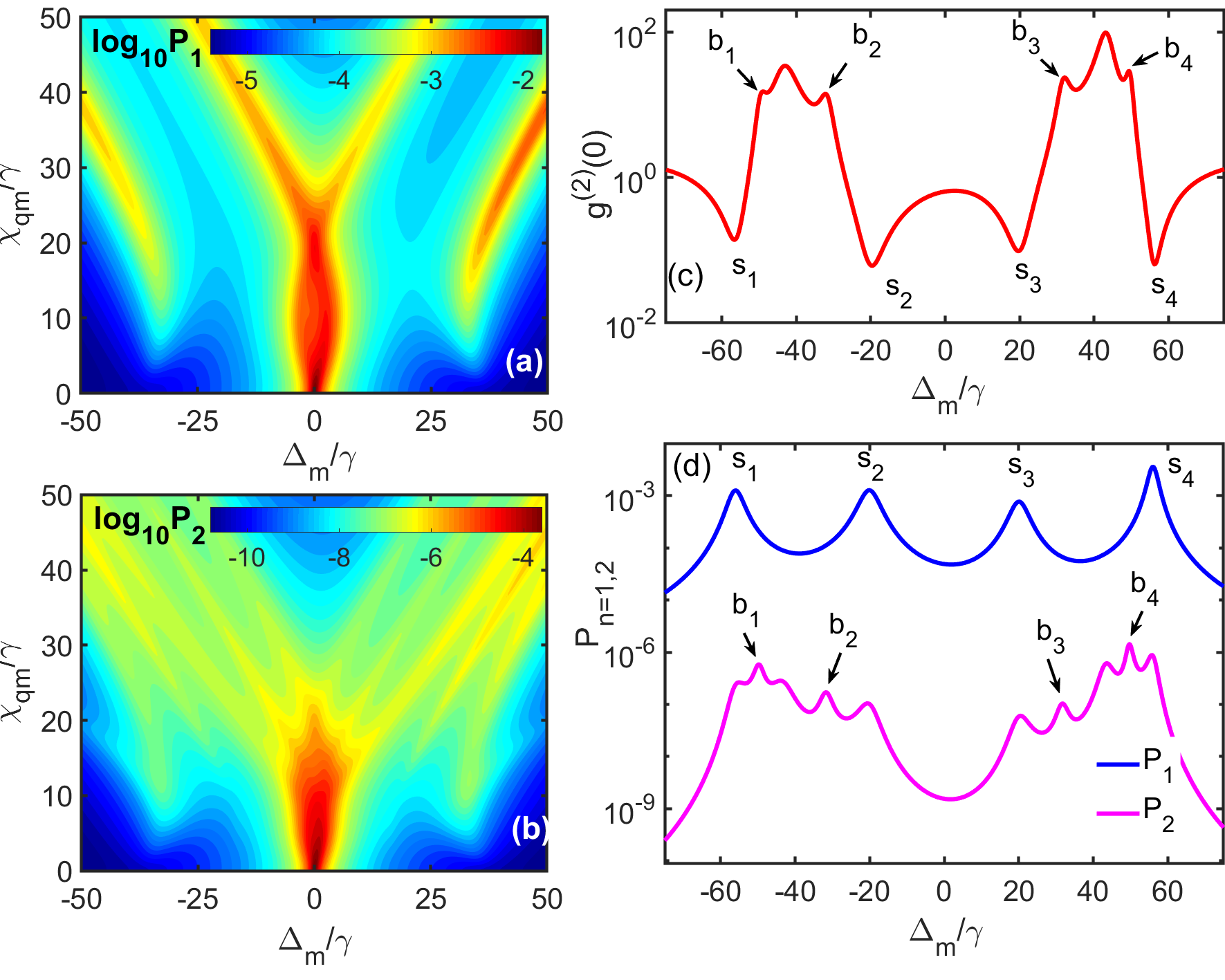}
\caption{The contour plot of the magnon-number distributions $\rm{log}_{10}P_{n=1,2}$ vary with the driving detuning $\Delta_{m}$ and the dispersive coupling strength $\chi_{qm}$.
The single-magnon-number distribution $\rm{log}_{10}P_{1}$ in (a) and two-magnon-number distribution $\rm{log}_{10}P_{2}$ in (b).
(c)-(d) The  second-order magnon coherence function $g^{(2)}(0)$ and the magnon-number distributions $P_{n=1,2}$ as functions of the detuning $\Delta_{m}$.
The dispersive coupling strength $\chi_{qm}/\gamma$ = 45 and the other parameters are the same as those in Fig. \ref{fig:2}.}
\label{fig:4}
\end{figure}
i.e., $g^{(2)}(0)\sim1$, indicating that the magnonic mode has neither observable bunching nor anti-bunching behavior.
With the increase of the dispersive coupling strength, significant magnonic bunching and anti-bunching phenomena can be advantageously observed.
To take two examples, when the qubit-magnon dispersive coupling strength $\chi_{qm}/\gamma = 20$ and $\chi_{qm}/\gamma = 40$, as shown in Figs. \ref{fig:2}(c) and (d), respectively, an optimal magnon blockade effect [$g^{(2)}(0) \rightarrow 0.04$] and a conspicuous magnon bunching effect [$g^{(2)}(0)\rightarrow 100$] are found under certain driving detuning conditions.
In these cases, the qubit-magnon dispersive coupling strengths are larger than the linewidths $\kappa_{m}$ of the magnonic mode and $\kappa_{q}$ of the qubit, thereby reaching the strong dispersive regime.
It should be noted that the impact of ambient thermal noise on the magnon blockade effect is not considered here (i.e., the thermal magnon and qubit number $m_{th} = n_{th} = 0$), which will be discussed in detail in Fig. \ref{fig:5}.
Additionally, we can observe a shift in the location where the magnon blockade occurs, as indicated by the brown arrows in Figs. \ref{fig:2}(c) and (d).
Physically, the qubit-magnon dispersive interaction leads to a shift in the qubit frequency, so the corresponding resonance conditions also need to change to satisfy the magnon resonance transition.
A clearer physical picture can be understood from the energy level diagram of the system.
As schematically shown in Fig. \ref{fig:1}(c), owing to this shift, the levels $|g,2\rangle$ and $|e,2\rangle$ have been moved out of the harmonic ladder.
And thus, starting from the zero magnon state, the magnon transition is confined to the levels $|g,0\rangle \rightarrow |g,1\rangle$, $|g,0\rangle \rightarrow |e,1\rangle$, $|e,0\rangle \rightarrow |g,1\rangle$ and $|e,0\rangle \rightarrow |e,1\rangle$, and any transition to the levels $|g,2\rangle$ and $|e,2\rangle$ is forbidden, as shown by the dotted lines in Fig. \ref{fig:1}(c).

In what follows, the magnon-number distributions $P_{n=1,2}$ vary with the driving detuning $\Delta_{m}$ and the dispersive coupling strength $\chi_{qm}$ is plotted in Fig. \ref{fig:4}.
As expected, the magnon-number distributions satisfy $P_{2}\ll P_{1}$ in the weak-driving case, and clearer results are shown in Figs. \ref{fig:4}(a) and (b).
In addition, we observe a high dependence of the magnon-number distributions $P_{n=1,2}$ on the dispersive coupling strength and the driving detuning, which shows an excellent agreement with the results discussed in Fig. \ref{fig:2}(a).
To give an example for detailed discussion, in the case of a strong dispersive coupling strength $\chi_{qm}/\gamma$ = 45, the second-order magnon coherence function $g^{(2)}(0)$ as
a function of the driving detuning $\Delta_{m}$ is shown in Fig. \ref{fig:4}(c).
We can see that there are some obvious extreme points $b_{n=1,2,3,4}$ and $s_{n=1,2,3,4}$ on the curve of $g^{(2)}(0)$, which correspond to the appearance of magnon bunching and anti-bunching, respectively.
More specifically, the locations of the four points $s_{n=1,2,3,4}$ in the curve of $g^{(2)}(0)$ correspond to the single-magnon resonant transitions in $P_{1}$.
Namely, the transitions $|g,0\rangle \rightarrow |g,1\rangle$ and $|e,0\rangle \rightarrow |e,1\rangle$ with the resonance conditions $\Delta_{m}/\gamma=\frac{1}{2\gamma}\left[\pm\sqrt{(\Delta_{q}+2\chi_{qm})^{2}+\Omega_{s}^{2}}
\mp\sqrt{\Delta_{q}^{2}+\Omega_{s}^{2}}\right]\approx\pm20$ correspond to the locations $s_{3}$ and $s_{2}$ in $P_{1}$, as shown by the blue curve in Fig. \ref{fig:4}(d).
The locations $s_{4}$ and $s_{1}$ in $P_{1}$ correspond to the transitions $|e,0\rangle \rightarrow |g,1\rangle$ and $|g,0\rangle \rightarrow |e,1\rangle$ with the resonance conditions $\Delta_{m}/\gamma=\frac{1}{2\gamma}\left[\pm\sqrt{(\Delta_{q}+2\chi_{qm})^{2}+\Omega_{s}^{2}}
\pm\sqrt{\Delta_{q}^{2}+\Omega_{s}^{2}}\right]\approx\pm56$, respectively.
Furthermore, the locations $b_{n=1,2,3,4}$ in the curve of $g^{(2)}(0)$ are associated with
the two-magnon resonant transitions in $P_{2}$, i.e., $|g,0\rangle \rightarrow |g(e),2\rangle$ and $|e,0\rangle \rightarrow |g(e),2\rangle$, as shown by the pink curve in Fig. \ref{fig:4}(d), with the corresponding resonance conditions $\Delta_{m}=\frac{1}{4}\left[\pm\sqrt{(\Delta_{q}+4\chi_{qm})^{2}+\Omega_{s}^{2}}
-\sqrt{\Delta_{q}^{2}+\Omega_{s}^{2}}\right]$ and $\Delta_{m}=\frac{1}{4}\left[\pm\sqrt{(\Delta_{q}+4\chi_{qm})^{2}+\Omega_{s}^{2}}
+\sqrt{\Delta_{q}^{2}+\Omega_{s}^{2}}\right]$, respectively.
Note that the other peaks in $P_{2}$ are induced by the single-magnon resonant transitions.
From the above discussion, we find that the optimal magnon blockade effect occurs at single-magnon resonant transitions, which is sensitively dependent on the frequency of the driving field.

In a realistic situation, the system is inevitably affected by its environment.
Therefore, evaluating the influence of ambient thermal noise on the magnon blockade effect is necessary.
Assuming that the system is connected to a high temperature thermal bath, 
both the qubit and magnonic modes are driven by thermal noise entering the system \cite{qubit5}.
The thermal noises are treated to be Gaussian random numbers with mean values ${{m}_{th}} =[{\rm{exp}}(\frac{\hbar\omega_{m}}{\rm{K_{\rm{B}}T}})-1]^{-1}$ (for magnonic mode) and ${{n}_{th}} =[{\rm{exp}}(\frac{\hbar\omega_{q}}{\rm{K_{\rm{B}}T}})-1]^{-1}$ (for qubit), with $\rm{T}$ the ambient temperature and $\rm{K_{B}}$ the Boltzmann constant.
The second-order magnon coherence function $g^{(2)}(0)$ varies with the thermal noise numbers of the qubit ($n_{th}$) and magnonic mode ($m_{th}$) is plotted in Fig. \ref{fig:5}(a).
Obviously, the ambient thermal noise has a significant impact on the magnon blockade effect. 
The difference is that the second-order magnon coherence function $g^{(2)}(0)$  is more sensitive to $m_{th}$ than $n_{th}$, as shown in Figs. \ref{fig:5}(b) and (c), respectively.
Physically, magnon thermal noise directly populates magnonic modes, while qubit noise indirectly affects the system via dephasing.
Furthermore, the critical thermal noise values of the qubit and magnon for observing magnon blockade are compared.
To be specific, when the thermal magnon number ${{m}_{th}} \sim 0.0035$, corresponding to the ambient temperature T $\sim$ 72 mK for the magnon frequency $\omega_{m}/2\pi = 8.5 \rm{GHz}$, the second-order magnon coherence function $g^{(2)}(0)=1$ [as indicated by the pink arrows in Fig. \ref{fig:5}(b)], showing that the magnon blockade effect has been destroyed, whereas, the qubit noise $n_{th}$ has negligible impact under such low-temperature condition.
Therefore, suppressing magnon thermal noise is crucial for practical implementation of the magnon blockade effect.
Under the current experimental conditions, the device can be placed within a dilution refrigerator with a base temperature of $\sim$ 46-48 mK \cite{qubit1,qubit3,qubit4,qubit5}, and thus, the aforementioned discussions about the magnon blockade effect in quantum magnonics are experimentally viable.

\section{CONCLUSION}
In summary, this study extends the investigation of the magnon blockade effect in the strong dispersive regime of quantum magnonics, deepening our understanding of dispersive interactions with a superconducting qubit.
We explored the dependence of the second-order magnon coherence function on the dispersive coupling, revealing the mechanisms that govern the magnon blockade phenomenon in this regime.
Compared with the previous protocols in the near-resonant regime, magnon blockade effect in the dispersive regime possesses several merits: First, the dispersive regime provides a more straightforward means of qubit readout and a measurement tool, which can be less complex than the near-resonant regime protocols \cite{advantages}.
Secondly, the dispersive regime can accommodate higher levels of the qubit in the virtual transitions that contribute to the dispersive shifts, leading to richer physics and potentially enhanced dispersive shifts \cite{qubit3}.
Third, from the fundamental aspect, magnon blockade effect in the dispersive regime is still new and remarkable for quantum magnonics.
Our results thus advance quantum magnonics by providing a robust, tunable platform for single-magnon manipulation with applications in quantum information and sensing.

\begin{acknowledgments}
This work was supported by the National Science Foundation (NSF) of China (Grants No. 12105047), Guangdong Provincial Quantum Science Strategic Initiative (GDZX2302002), National Key Research and Development Program of China (2022YFB4601103), Guangdong Basic and Applied Basic Research Foundation (Grant No. 2022A1515010446, 2021A1515110612).
\end{acknowledgments}


\begin{thebibliography}{50}

\bibitem{mode2} Y. Tabuchi, S. Ishino, A. Noguchi, T. Ishikawa, R. Yamazaki, K. Usami, and Y. Nakamura, Quantum magnonics: The magnon meets the superconducting qubit, C.R. Phys. 17, 729 (2016).

\bibitem{Hybrid} L.-Q. Dany, T. Yutaka, G. Arnaud, U. Koji, and N. Yasunobu, Hybrid quantum systems based on magnonics, Appl. Phys. Express 12, 070101 (2019).

\bibitem{quantum1} H.-Y. Yuan , Y.-S. Cao, A. Kamra, R. A. Duine, P. Yan, Quantum magnonics: When magnon spintronics meets quantum information science, Phys. Rep. 965, 1 (2022).

\bibitem{quantum2} B.Z. Rameshti, S.V. Kusminskiy, J.A. Haigh, K. Usami, D. Lachance-Quirion, Y. Nakamura, C.-M. Hu, H.X. Tang, G.E.W. Bauer, Y.M. Blanter, Cavity magnonics, Phys. Rep. \textbf{979}, 1 (2022).

\bibitem{quantum3} X. Zuo, Z. Y. Fan, H. Qian, M. S. Ding, H. Tan, H. Xiong, and J. Li, Cavity magnomechanics: from classical to quantum, New J. Phys. 26, 031201 (2024).
\bibitem{FOR3} K. Wang, Y. P. Gao, R. Jiao, and C. Wang, Recent progress on optomagnetic coupling and optical manipulation based on cavity-optomagnonics, Front. Phys. 17, 42201 (2022).
\bibitem{Xiong} H. Xiong, Center-of-mass magnomechanics beyond magnetostrictive limits, Sci. China-Phys. Mech. Astron. 68, 250313 (2025).

\bibitem{entanglement1} J. Li, S.-Y. Zhu, and G.S. Agarwal, Magnon-photon-phonon entanglement in cavity magnomechanics, Phys. Rev. Lett. 121, 203601 (2018).

\bibitem{entanglement2} Z. Zhang, M.O. Scully, and G.S. Agarwal, Quantum entanglement between two magnon modes via Kerr nonlinearity driven far from equilibrium, Phys. Rev. Res. 1, 023021 (2019).

\bibitem{entanglement3} M. Yu, H. Shen, and J. Li, Magnetostrictively induced stationary entanglement between two microwave fields, Phys. Rev. Lett. 124, 213604 (2020).

\bibitem{entanglement4} V. Azimi Mousolou, Y. Liu, A. Bergman, A. Delin, O. Eriksson, M. Pereiro, D. Thonig, E. Sj\"{o}qvist, Magnon-magnon entanglement and its quantification via a microwave cavity, Phys. Rev. B 104, 224302 (2021).

\bibitem{entanglement5} Y.-l. Ren, J.-k. Xie, X.-k. Li, S.-l. Ma, and F.-l. Li, Long-range generation of a magnon-magnon entangled state, Phys. Rev. B 105, 094422 (2022).
\bibitem{FOR1} J. M. Li and S. M. Fei, Quantum entanglement generation on magnons assisted with microwave cavities coupled to a superconducting qubit, Front. Phys. 18, 41301 (2023).

\bibitem{Squeezed} J. Zhao, A. V. Bragas, D. J. Lockwood, and R. Merlin, Magnon Squeezing in an Antiferromagnet: Reducing the Spin Noise below the Standard Quantum Limit, Phys. Rev. Lett. 93, 107203 (2004).

\bibitem{Squeezed1} J. Li, S.-Y. Zhu, G.S. Agarwal, Squeezed states of magnons and phonons in cavity magnomechanics, Phys. Rev. A 99, 021801, (2019).


\bibitem{cat} S. Sharma, V.A.S.V. Bittencourt, A.D. Karenowska, and S.V. Kusminskiy, Spin cat states in ferromagnetic insulators, Phys. Rev. B 103, L100403 (2021).

\bibitem{cat1} F.-X. Sun, S.-S. Zheng, Y. Xiao, Q. Gong, Q. He, and K. Xia, Remote generation of magnon Schr\"{o}dinger cat state via magnon-photon entanglement, Phys. Rev. Lett. 127, 087203 (2021).


\bibitem{cat2} M. Kounalakis, G.E.W. Bauer, and Y.M. Blanter, Analog Quantum Control of Magnonic Cat States on a Chip by a Superconducting Qubit, Phys. Rev. Lett. 129, 037205 (2022).

\bibitem{cat3} S. He, X. Xin, F.-Y. Zhang, and C. Li, Generation of a Schr\"{o}dinger cat state in a hybrid ferromagnet-superconductor system, Phys. Rev. A 107, 023709 (2023).


\bibitem{cat4} D.-W. Liu, Y. Wu, and L.-G. Si, Magnon cat states induced by photon parametric coupling, Phys. Rev. Applied 21, 044018 (2024).


\bibitem{Einstein} S.O. Demokritov, V.E. Demidov, O. Dzyapko, G.A. Melkov, A.A. Serga, B. Hillebrands, and A.N. Slavin, Bose-Einstein condensation of quasi-equilibrium magnons at room temperature under pumping, Nature 443, 430, (2006).


\bibitem{Einstein1} T. Giamarchi, C. R\"{u}egg, and O. Tchernyshyov, Bose–Einstein condensation in magnetic insulators, Nat. Phys. 4, 198 (2008).


\bibitem{Einstein2} M. Schneider, T. Br\"{a}cher, D. Breitbach, V. Lauer, P. Pirro, D.A. Bozhko, H. Yu. M. Shmarova, B. Heinz, Q. Wang, T. Meyer, F. Heussner, S. Keller, E.T. Papaioannou, B. L\"{a}gel, T. L\"{o}ber, C. Dubs, A.N. Slavin, V.S. Tiberkevich, A.A. Serga, B. Hillebrands, and A.V. Chumak, Bose–Einstein condensation of quasiparticles by rapid cooling. Nat. Nanotechnol. 15, 457 (2020).


\bibitem{Einstein3} B. Divinskiy, H. Merbouche, V.E. Demidov, K.O. Nikolaev, L. Soumah, D. Gou\'{e}r\'{e}, R. Lebrun, V. Cros, J.B. Youssef, P. Bortolotti, A. Anane, and S.O. Demokritov, Evidence for spin current driven bose-Einstein condensation of magnons, Nat. Commun. 12, 6541 (2021).


\bibitem{qubit1} Y. Tabuchi, S. Ishino, A. Noguchi, T. Ishikawa, R. Yamazaki, K. Usami, and Y. Nakamura, Coherent coupling between a ferromagnetic magnon and a superconducting qubit, Science \textbf{349}, 405 (2015).

\bibitem{qubit3} D. Lachance-Quirion, Y. Tabuchi, S. Ishino, A. Noguchi, T. Ishikawa, R. Yamazaki, and Y. Nakamura, Resolving quanta of collective spin excitations in a millimeter-sized ferromagnet, Sci. Adv. \textbf{3}, e1603150 (2017).

\bibitem{qubit4} S. P. Wolski, D. Lachance-Quirion, Y. Tabuchi, S. Kono, A. Noguchi, K. Usami, and Y. Nakamura, Dissipationbased quantum sensing of magnons with a superconducting qubit, Phys. Rev. Lett. 125, 117701 (2020).
\bibitem{qubit5} D. Lachance-Quirion, S. P. Wolski, Y. Tabuchi, S. Kono, K. Usami, and Y. Nakamura, Entanglement-based single-shot detection of a single magnon with a superconducting qubit, Science 367, 425 (2020).
\bibitem{Absorption} Z.-X. Liu, H. Xiong, M.-Y. Wu, and Y.-Q. Li, Absorption of magnons in dispersively coupled hybrid quantum systems, Phys. Rev. A 103, 063702 (2021).

\bibitem{non-demolition} P. Grangier, J. A. Levenson, and J. P. Poizat, Quantum non-demolition measurements in optics, Nature, 396, 537–542 (1998).

\bibitem{network1} P. Pirro, V.I. Vasyuchka, A.A. Serga, and B. Hillebrands, Advances in coherent magnonics, Nat Rev Mater 6, 1114-1135 (2021).

\bibitem{network2} Y.-P. Wang, G.-Q. Zhang, D. Xu, T.-F. Li, S.-Y. Zhu, J.-S. Tsai, and J.-Q. You, Quantum Simulation of the Fermion-Boson Composite Quasi-Particles with a Driven Qubit-Magnon Hybrid Quantum System, arXiv:1903.12498.

\bibitem{network} J. Li, Y.-P. Wang, W.-J. Wu, S.-Y. Zhu, and J.Q. You, Quantum Network with Magnonic and Mechanical Nodes, PRX Quantum 2, 040344 (2021).

\bibitem{network66} Z.-X. Liu, Y.-Q. Li, Optomagnonic frequency combs, Photon. Res. 10, 467595 (2022).

\bibitem{network666} Z.-X. Liu, Dissipative coupling induced UWB magnonic frequency comb generation, Appl. Phys. Lett. 124, 032403 (2024).


\bibitem{network3} D. Xu, X.-K. Gu, H.-K. Li, Y.-C. Weng, Y.-P. Wang, J. Li, H. Wang, S.-Y. Zhu, and J. Q. You, Quantum control of a single magnon in a macroscopic spin system, Phys. Rev. Lett. 130, 193603 (2023).

\bibitem{network4} B. Flebus, S. M. Rezende, D. Grundler, and A. Barman, Recent advances in magnonics, J. Appl. Phys. 133, 160401 (2023).

\bibitem{network5} Q. Wang, G. Csaba, R. Verba, A. V. Chumak, and P. Pirro, Nanoscale magnonic networks, Phys. Rev. Applied 21, 040503 (2024).
\bibitem{network6} Z. X. Liu and H. Xong, Ultra-slow spin waves propagation based on skyrmion breathing, New J. Phys. 25, 103052 (2023).

\bibitem{Coulomb1} T. A. Fulton and G. J. Dolan, Observation of single-electron charging effects in small tunnel junctions, Phys. Rev. Lett. 59, 109 (1987).
\bibitem{Coulomb2} M. Kastner, The single-electron transistor, Rev. Mod. Phys. 64, 849 (1992).


\bibitem{blockade1} Z.-X. Liu, H. Xiong, and Y. Wu, Magnon blockade in a hybrid ferromagnet-superconductor quantum system, Phys. Rev. B 100, 134421 (2019).

\bibitem{blockade2} H.Y. Yuan and R.A. Duine, Magnon antibunching in a nanomagnet, Phys. Rev. B 102, 100402 (2020).

\bibitem{blockade3} J.-K. Xie, S.-L. Ma, and F.-L. Li, Quantum-interference-enhanced magnon blockade in an yttrium-iron-garnet sphere coupled to superconducting circuits, Phys. Rev. A 101, 042331 (2020).

\bibitem{blockade4} L. Wang, Z.-X. Yang, Y.-M. Liu, C.-H. Bai, D.-Y. Wang, S. Zhang, and H.-F. Wang, Magnon blockade in a PT-symmetric-like cavity magnomechanical system, Ann. Der Phys. 532, 2000028 (2020).

\bibitem{blockade5} C. Zhao, X. Li, S. Chao, R. Peng, C. Li, and L. Zhou, Simultaneous blockade of a photon, phonon, and magnon induced by a two-level atom, Phys. Rev. A 101, 063838 (2020).

\bibitem{blockade6} Y.-J. Xu, T.-L. Yang, L. Lin, and J. Song, Conventional and unconventional magnon blockades in a qubit-magnon hybrid quantum system, J. Opt. Soc. Am. B 38, 876 (2021).

\bibitem{blockade7} Y. Wang, W. Xiong, Z. Xu, G.-Q. Zhang, and J.-Q. You, Dissipation induced nonreciprocal magnon blockade in a magnon based hybrid system, Sci. China Phys. Mech. Astron. 65, 260314 (2022).


\bibitem{blockade8} F. Wang, C. Gou, J. Xu, and C. Gong, Hybrid magnon-atom entanglement and magnon blockade via quantum interference Phys. Rev. A 106, 013705 (2022).


\bibitem{blockade9} Z.-Y. Jin and J. Jing, Magnon blockade in magnon-qubit systems, Phys. Rev. A 108, 053702 (2023).

\bibitem{Wang11} W. Zhang, S. Liu, S. Zhang, and H.-F. Wang, Magnon blockade induced by parametric amplification, Phys. Rev. A 109, 043712 (2024).
\bibitem{blockade10} K.-W. Huang, X. Wang, Q.-Y. Qiu, and H. Xiong, Nonreciprocal magnon blockade via the Barnett effect, Opt. Lett. 49, 758--761 (2024).

\bibitem{blockade11} H.-Q. Zhang, S.-S. Chu, J.-S. Zhang, W.-X. Zhong, and G.-L. Cheng, Nonreciprocal magnon blockade based on nonlinear effects, Opt. Lett. 49, 2009-2012 (2024).

\bibitem{FOR2} C. Kong, J. B. Liu, and H. Xiong, Nonreciprocal microwave transmission under the joint mechanism of phase modulation and magnon Kerr nonlinearity effect, Front. Phys. 18, 12501 (2023).

\bibitem{magnon} C. Kittel, On the theory of ferromagnetic resonance absorption, Phys. Rev. 73, 155 (1948).

\bibitem{YIG} A.A. Serga, A.V. Chumak, and B. Hillebrands, YIG magnonics, J. Phys. D \textbf{43}, 264002 (2010).
    
\bibitem{ac-Stark} A. Blais, R.-S, Huang, A. Wallraff, S. M. Girvin, and R. J. Schoelkopf, Cavity quantum electrodynamics for superconducting electrical circuits: An architecture for quantum computation, Phys. Rev. A 69, 062320 (2004).
\bibitem{ac-Stark1} J. Gambetta, A. Blais, D. I. Schuster, A. Wallraff, L. Frunzio, J. Majer, M. H. Devoret, S. M. Girvin, and R. J. Schoelkopf, Qubit-photon interactions in a cavity: Measurement-induced dephasing and number splitting, Phys. Rev. A 74, 042318 (2006).

\bibitem{Kerr} G. Zhang, Y. Wang, and J. You, Theory of the magnon Kerr effect in cavity magnonics, Sci. China Phys. Mech. Astron. 62, 987511 (2019).

\bibitem{HP} T. Holstein and H. Primakoff, Phys. Rev. 58, 1098 (1940).

\bibitem{Low} J. Peng, Z. X. Liu, Y. F. Yu, and H. Xiong, Cavity magnomechanical chaos, Phys. Rev. A 110, 053704 (2024).


\bibitem{mast1} D. Walls and G. Milburn, Quantum optics (Spinger-Verlag, Berlin, 1994).


\bibitem{QuTiP} J. R. Johansson, P. D. Nation, F. Nori, QuTiP 2: A Python framework for the dynamics of open quantum systems. Comput. Phys. Commun. 184, 1234 (2013).


\bibitem{block} A. Imamoglu, H. Schmidt, G. Woods, and M. Deutsch, Strongly Interacting Photons in a Nonlinear Cavity, Phys. Rev. Lett. 79, 8 (1997).


\bibitem{advantages} G. Zhu, D. G. Ferguson, V. E. Manucharyan, and J. Koch, Circuit QED with fluxonium qubits: Theory of the dispersive regime, Phys. Rev. B 87, 024510 (2013).


\end{thebibliography}
\end{document}